\documentclass[10pt]{amsart}
\usepackage{amssymb}
\usepackage{enumerate}
\usepackage[utf8]{inputenc}
\usepackage{graphicx}
\usepackage{hyperref}
\usepackage{mathrsfs}
\usepackage{float}
\usepackage[export]{adjustbox}
\newcommand \intl[4]{ \int\limits_{#1}^{#2}{#3}{#4}      }   % integral with limits
\newcommand{\diag}{\operatorname{diag}}
\begin{document}
\begin{abstract}
We present a time-dependent solution of the Maxwell equations in the Einstein universe, whose electric and magnetic fields, as seen by the stationary observers, are aligned with the Clifford parallels of the $3$-sphere $S^3$. The conformal equivalence between Minkowski's spacetime and (a region of) the Einstein cylinder is then exploited in order to obtain a knotted, finite energy, radiating solution of the Maxwell equations in flat spacetime. We also discuss similar electromagnetic fields in expanding closed Friedmann models, and compute the matter content of such configurations.
\end{abstract}
\title{On a remarkable electromagnetic field in the Einstein Universe}
\author{Jaros\l{}aw Kopi\'nski and Jos\'e Nat\'ario}
\address{
Institute of Theoretical Physics, Faculty of Physics, University of Warsaw, Pasteura 5, 02-093 Warsaw, Poland}
\address{
Center for Mathematical Analysis, Geometry and Dynamical Systems, Mathematics Department, Instituto Superior T\'ecnico, Universidade de Lisboa, Portugal}
\thanks{Partially funded by FCT/Portugal through projects UID/MAT/04459/2013 and PTDC/MAT-ANA/1275/2014.}
\maketitle
\section*{Introduction}
In the early days of general relativity there was a common belief that the Universe had to be eternal and unchanging. Following this concept, Albert Einstein introduced the first cosmological model, subsequently named after him. It consisted of a compact spatial hypersurface with positive curvature, the $3$-sphere $S^3$, which did not change under the flow of time. In order to overcome the attraction force of matter and obtain a static configuration, a new term had to be introduced in the field equations -- the famous cosmological constant. Although numerous observations have since excluded this kind of static models in favor of expanding ones, the Einstein universe is still a very important explicit solution of the Einstein equations, partly because of the conformal equivalence between (a region of) this model and Minkowski's spacetime (the standard background for quantum field theories). For example, since Maxwell's equations are conformally invariant, one can think of any configuration of electromagnetic fields in the Einstein universe as a configuration in flat spacetime.
\\
\indent
The Einstein universe is a particular case of, and conformal to, the closed Friedmann-Lema\^\i tre-Robertson-Walker (FLRW) models, where the assumption of staticity is relaxed by introducing a scale factor that allows the universe to expand or contract. Using this relation, one can extend solutions of Maxwell's equations in the Einstein universe to the closed FLRW models. In general, when one allows the spatial volume to change in time, the energy of the electromagnetic configuration will also change according to an appropriate power of the scale factor.
\\
\indent
The common denominator of the models described above is the $3$-sphere $S^3$. This manifold is an important example of a non-trivial fibre bundle, given by the Hopf fibration, whose base is the usual $2$-sphere $S^2$, and whose fibres are circles $S^1$ (the Clifford parallels). It is natural to look for solutions of the Maxwell equations whose electric and magnetic fields, as measured by the static observers, are aligned with these fibres. Such ansatz will be used in the first section to find a remarkable solution of Maxwell's equations in the Einstein universe, which will then be interpreted as a knotted, finite energy, radiating electromagnetic field in Minkowski's spacetime. The extension of this solution to closed FLRW models will be carried out in the second section. Finally, the matter distribution that must be added to obtain a self-consistent solution of the Einstein-Maxwell equations will be determined in the third section.
\\
\indent
We follow the conventions of \cite{MTW, wald}; in particular, we use the Einstein summation convention and a system of units in which $c = G = 1$.

\section{Electromagnetic field in the Einstein universe}
\begin{figure}[h]
\centering
\includegraphics[scale=0.3]{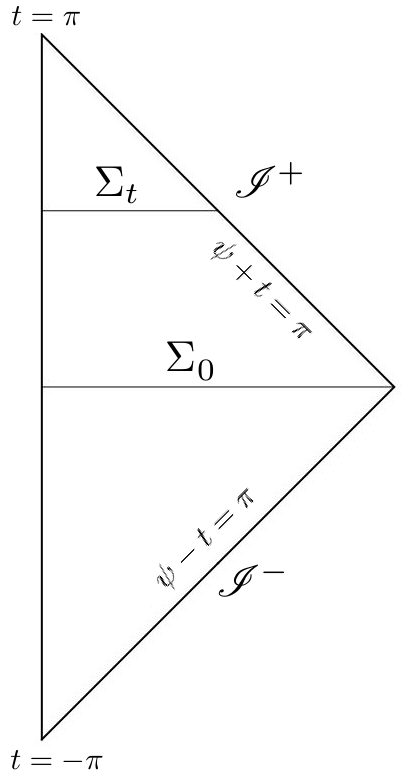}
\caption{Compactified Minkowski spacetime with two dimensions suppressed. Each point represents a sphere (possibly degenerated).}
\label{fig: figure3}
\end{figure}

We are interested in finding solutions of the Maxwell equations in the cylinder $\mathcal{E}=\mathbb{R} \times S^3$ with the standard Lorentzian metric. This manifold represents a static universe with positive cosmological constant $\Lambda$, called the Einstein universe. By choosing the radius of this universe as our unit length we can assume, without any loss of generality, that $\mathcal{E}$ is the product of $\mathbb{R}$ by the unit sphere.
\\
\indent
Since $S^3\cong SU(2)$ is a Lie group, so is $\mathcal{E}$. We can introduce the following left-invariant orthonormal tetrad of one-forms on this manifold: Take $\theta^0 = \mathrm{d}t$ to be the standard element of the  holonomic basis on the cylinder, defining the cross section foliation. On each leaf, an orthonormal triad $\{\theta^1,\theta^2,\theta^3 \}$ is chosen in such a way that\footnote{Because we have chosen an orthonormal triad, we do not have to worry about the position of the spatial indices $i,j,k,\ldots$ being subscript or superscript.}
\begin{equation} \label{eq: mcr}
\mathrm{d} \theta^i = -\varepsilon_{ijk} \theta^j \wedge \theta^k,
\end{equation}
thus reproducing the $\mathfrak{su}(2)$ Lie algebra structure. In terms of the dual orthonormal vector tetrad, the choice of $\theta^0$ and relation (\ref{eq: mcr}) can be written as
\begin{equation}
X_0 = \partial_t, \quad [X_i,X_j] =2 \varepsilon_{ijk} X_k.
\end{equation}
\indent
We look for a solution of the Maxwell equations in vacuum,
\begin{equation} \label{eq:Maxwell}
dF = d \star F = 0,
\end{equation}
with the Faraday two-form 
\begin{equation}
F = E^i \theta^i \wedge \mathrm{d}t + \frac{1}{2}B^i \varepsilon_{ijk} \theta^j \wedge \theta^k.
\end{equation}
Written in this basis, they are
\begin{equation} \label{eq:MaE}
\begin{cases} 
X_i(E^i)=X_i(B^i)=0, \\ 
\dot{B^i} -2 E^i + \varepsilon_{ijk} X_j(E^k)=0,\\ 
\dot{E^i} +2 B^i - \varepsilon_{ijk} X_j(B^k)=0.
 \end{cases}
\end{equation}
The simplest nontrivial solution can be obtained from the assumption that the components $E^i$ and $B^i$ of the electric and magnetic fields $E$ and $B$ measured by the stationary observers depend only on time. Using that ansatz, we arrive at the relations
\begin{equation}  \label{eq: maxtime}
\begin{cases} 
\dot{B^i} = 2 E^i, \\ 
\dot{E^i} = - 2 B^i .
 \end{cases}
\end{equation}
A solution of system (\ref{eq: maxtime}) will include trigonometric functions with different initial phases and field strength factors. We choose the simplest one for the conformal analysis,
\begin{equation} \label{eq: maxsol}
 E=E_0\cos(2t)X_1, \quad  B = E_0\sin(2t)X_1,
\end{equation}
 because the results will not change qualitatively under more general assumptions\footnote{The property that all field lines (as seen by the stationary observers) are closed is unstable: any generic electromagnetic perturbation will destroy this feature. Moreover, such small perturbations will not decay away in time, as they would in Minkowski's spacetime, since $S^3$ is compact.}.
\\
\indent
Recall that Minkowski's spacetime is conformal to an open region of the Einstein universe \cite{wald, HE95}. More precisely, the Minkowski metric $\widetilde{g}$ can be written as
\begin{equation}
\widetilde{g}_{ab} = \Omega^{-2} g_{ab},\quad  \Omega =2 \cos\left( \frac{t-\psi}{2}\right) \cos\left( \frac{t+\psi}{2}\right),
\end{equation}
where $(\psi, \theta, \varphi)$ are the standard hyperspherical coordinates on the 3-sphere. In these coordinates, Minkowski's spacetime corresponds to the region $0 \leq |t| + \psi < \pi$ (Fig. \ref{fig: figure3}).

Since Maxwell's equations are conformally invariant, we can interpret \eqref{eq: maxsol} as an electromagnetic field in flat spacetime\footnote{See \cite{Bicak1, Bicak2} for a similar contruction exploiting the conformal relation between Minkowski's and de Sitter's spacetimes.}. The energy will of course change under the conformal transformation. Its value will be given by the integral expression
\begin{equation}
\widetilde{H}(t)=\intl{\widetilde{\Sigma}_{t}}{}{\widetilde{T}(\widetilde{X}_0\widetilde{X}_0)}{{\mathrm{d}(\widetilde{vol})}},
\end{equation}
where
\begin{equation} \label{eq: setef}
\widetilde{T}_{ab} = \widetilde{F}_{ac} \widetilde{F}_{b}^{\,\,c} - \frac14 \widetilde{g}_{ab}\widetilde{F}_{cd}\widetilde{F}^{cd}
\end{equation}
is the energy-momentum tensor of $\widetilde{F}$, $\widetilde{X}_0 = \Omega X_0$ is $X_0$ normalized by the Minkowski metric $\widetilde{g}$, and $\widetilde{\Sigma}_t$ is the spacelike hypersurface corresponding to the hyperspherical cap $\Sigma_t$ of constant $t$ (Fig.~\ref{fig: figure3}). Note that $\widetilde{\Sigma}_t$ is not a Cauchy surface for Minkowski's spacetime unless $t=0$, as it approaches $\mathscr{I^-}$ for $t<0$ and $\mathscr{I^+}$ for $t>0$.

\begin{figure}[h]
\centering
\includegraphics[scale=0.45]{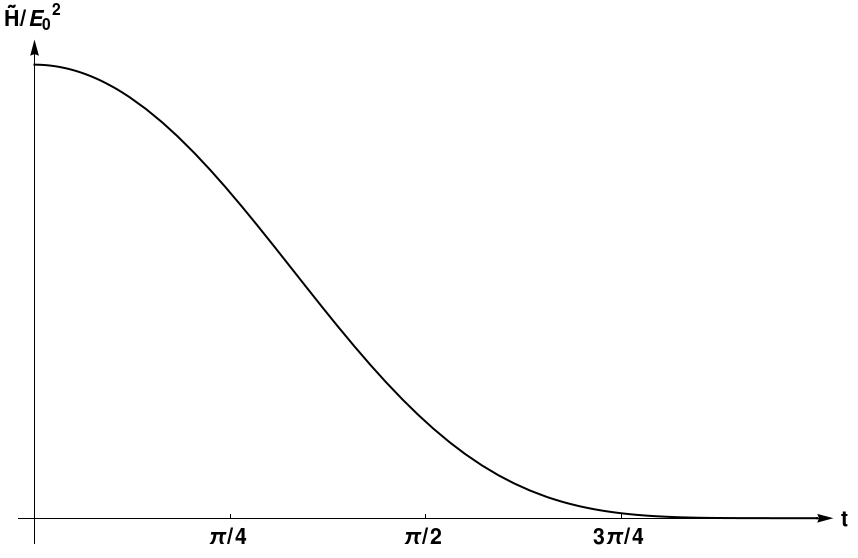}
\caption{Energy of the electromagnetic field in the Minkowski spacetime.}
\label{fig: figure1}
\end{figure}

Using $\widetilde{T}_{ab} = \Omega^2{T}_{ab}$ and ${\mathrm{d}(\widetilde{vol})} = \Omega^{-3}\mathrm{d}(vol)$, we can reexpress the energy as an integral over the hyperspherical cap $\Sigma_t$,
\begin{equation}
\widetilde{H}(t)=\int_{\Sigma_t}{\Omega \ T(X_0,X_0)}{\mathrm{d}(vol)}.
\end{equation}
It is straightforward to see that this quantity is finite. Moreover, it is a decreasing function of the parameter $t>0$, which indicates that the solution under consideration describes radiation fields. This can be seen by looking at the time dependence of the energy (Fig.~\ref{fig: figure1}.), given, for $t>0$, by the formula
\begin{equation}
\widetilde{H}(t) =\frac{{E_0}^2}{12} \left(9 \pi  \sin t+\pi  \sin 3 t+12 \pi ^2 \cos t-12 \pi  t \cos t \right).
\end{equation}
\\
\indent 
The electromagnetic field \eqref{eq: maxsol} is related to the simplest initial conditions given by Ra\~nada in \cite[Section 2]{ranada2} (see also \cite{ranada1, ranada3}), where only the initial magnetic field is nonzero. In fact, the phase-shifted solution (with $E_0=1$)
\begin{equation}
E=-\sin(2t)X_1, \quad  B = \cos(2t)X_1
\end{equation}
is the Cauchy development of initial conditions of that type. Due to the fact that it is aligned along the Clifford parallels, the Hopf index $n$ is equal to one in our case. This fact can be confirmed using an integral expression for this quantity, namely
\begin{equation}
n_{t=0}=-\frac{1}{ \pi^2}\intl{\{t=0\}}{}{A \wedge F} \  = \  \frac{1}{ 2\pi^2}\intl{S^3}{}{\theta^1 \wedge \theta^2 \wedge \theta^3} \  = \ 1, 
\end{equation}
where $F = \mathrm{d}A$.
% \indent
% In order to investigate the existence of more general solutions of the Maxwell equations we turn into the beginning and write them down as equations considering potential one-form $A$,
% \begin{equation}
% \begin{cases} 
% \mathrm{d} \mathrm{d} A =0, \\ 
% \mathrm{d} \star \mathrm{d} A=0.\\ 
%  \end{cases}
% \end{equation}
% One can see that the first one is just a claim that $F=\mathrm{d}A$ is a closed one-form. As for the second one, it is equivalent to the Laplace equation, for which maximum principle asserts constant solutions on $S^3$. Having said that, we state that the only possible expressions for fields $E$ and $B$ are just time dependent and can be obtained from system (\ref{eq: maxtime}).
%
\section{Expanding models}
In this section we turn our attention to the expanding models. Consider a metric of the form
\begin{equation} \label{eq: flrwc}
g=-\mathrm{d}t^2 + a(t) \gamma_{S^3}, 
\end{equation}
where $\gamma_{S^3}$ is the standard metric of $S^3$ with unit radius; this corresponds to a closed FLRW model, widely used in modern cosmology \cite{MTW}.
\\
\indent
In analogy with the previous case, we introduce the tetrad $\{\bar{\theta}^0,\bar{\theta}^1,\bar{\theta}^2,\bar{\theta}^3\}$,
\begin{equation}
\bar{\theta}^0 = \theta^0 = \mathrm{d}t, \quad \bar{\theta}^i = a(t) \theta^i,
\end{equation}
and the dual vector basis $\{\bar{X}_0,\bar{X}_1,\bar{X}_2,\bar{X}_3 \}$. 
\\
\indent
We are looking for a solution of the Maxwell equations similar to \eqref{eq: maxsol}, namely with a Faraday tensor of the form
\begin{equation}
F=E^1 \bar{\theta}^1 \wedge \bar{\theta}^0 + B^1\bar{\theta}^2 \wedge \bar{\theta}^3,
\end{equation}
where $E^1$ and $B^1$ depend only on time. Substituting into the vacuum Maxwell equations, \eqref{eq:Maxwell}, we have
\begin{equation} \label{eq: maxsoltime}
\begin{cases} 
\frac{d}{dt} \left( a^2 B^1 \right) = 2 a E^1, \\
\frac{d}{dt} \left( a^2 E^1 \right) = - 2 a B^1,
 \end{cases}
\end{equation}
yielding
\[
E^1 = \frac{E_0}{a^2}\cos\left(2\int\frac{dt}{a}\right), \quad  B^1 = \frac{E_0}{a^2}\sin\left(2\int\frac{dt}{a}\right).
\]
The Einstein universe case corresponds to $a\equiv 1$. Note that the $T_{00}$ component of the electromagnetic stress-energy tensor is
\begin{equation}
\frac{1}{2}\left( \left(E^1\right)^2 + \left(B^1\right)^2 \right) = \frac{{E_0}^2}{2a^4},
\end{equation}
indicating that the energy density of the solution is decaying with the fourth power of the radius of the universe.
\\
% \indent
% We can see that the full solution of (\ref{eq: maxsoltime}) can be obtained by adding another equation for scale factor $a$, which comes from the Einstein field equations. 
\section{Matter content}
The solutions of Maxwell's equations discussed in the last sections were derived under the assumption of a fixed background metric, satisfying the Einstein field equations
\begin{equation} 
G_{ab} +\Lambda g_{ab} = 8 \pi T^{\text{tot}}_{ab},
\end{equation}
where $G_{ab}$ is the Einstein tensor, $g_{ab}$ is the metric and $T^{\text{tot}}_{ab}$ is the total stress-energy tensor. One can ask what kind of matter has to be present for this configuration to be self-consistent. In order to answer this question, we consider the decomposition of the full stress-energy tensor into the parts coming from the electromagnetic field and matter,
\begin{equation}
T^{\text{tot}}_{ab}=T_{ab}+T^{\text{mat}}_{ab}.
\end{equation}
We already used the expression for the time-time component of $T_{ab}$ when computing the energy of the solution (\ref{eq: maxsol}); the full expression for this tensor is given by (\ref{eq: setef}).
It is straightforward to see that it has diagonal form for any solution of the system (\ref{eq: maxsoltime}), $T_{ab} = \diag(D, -D, D,D)$, where in general $D=D(t)$. 
%Using that, we obtain an expression for $T^{\text{mat}}$,
%\begin{equation} \label{eq: setm}
%T^{\text{mat}} = \frac{1}{8 \pi}(G + \Lambda g) - T.
%\end{equation}

\indent
Let us start from the assumption that matter can be described by a density $\rho$ and a principal pressure $p$ aligned with $X_1$,
\begin{equation} \label{eq: setmex}
T^{\text{mat}}_{00}=\rho, \quad T^{\text{mat}}_{11}=p. 
\end{equation}
A possible interpretation of this anisotropic pressure can be given by superimposing the stress-energy tensors of two pressureless fluids with the same mass density $\mu$, moving in opposite directions along $X_1$, with the same velocity with respect to the fundamental observers:
\begin{align*}
&
\begin{bmatrix}
\mu U^0 U^0 & \mu U^0 U^1 & 0 & 0\\
\mu U^0 U^1 & \mu U^1 U^1 & 0 & 0\\
0 & 0 & 0 & 0\\
0 & 0 & 0 & 0\\
 \end{bmatrix}
 +
 \begin{bmatrix}
\mu U^0 U^0 & -\mu U^0 U^1 & 0 & 0\\
-\mu U^0 U^1 & \mu U^1 U^1 & 0 & 0\\
0 & 0 & 0 & 0\\
0 & 0 & 0 & 0\\
 \end{bmatrix}
 \\
 &
 =
 \begin{bmatrix}
2\mu U^0 U^0 & 0 & 0 & 0\\
0 & 2\mu U^1 U^1 & 0 & 0\\
0 & 0 & 0 & 0\\
0 & 0 & 0 & 0\\
 \end{bmatrix}
 =
 \begin{bmatrix}
\rho & 0 & 0 & 0\\
0 & p & 0 & 0\\
0 & 0 & 0 & 0\\
0 & 0 & 0 & 0\\
 \end{bmatrix}
 .
\end{align*}
Another possible interpretation can be given by considering cosmic strings aligned with the Clifford parallels, expanding and contracting with space while preserving the structure of the Hopf fibration.

\begin{figure}[h]
\includegraphics[scale=0.4]{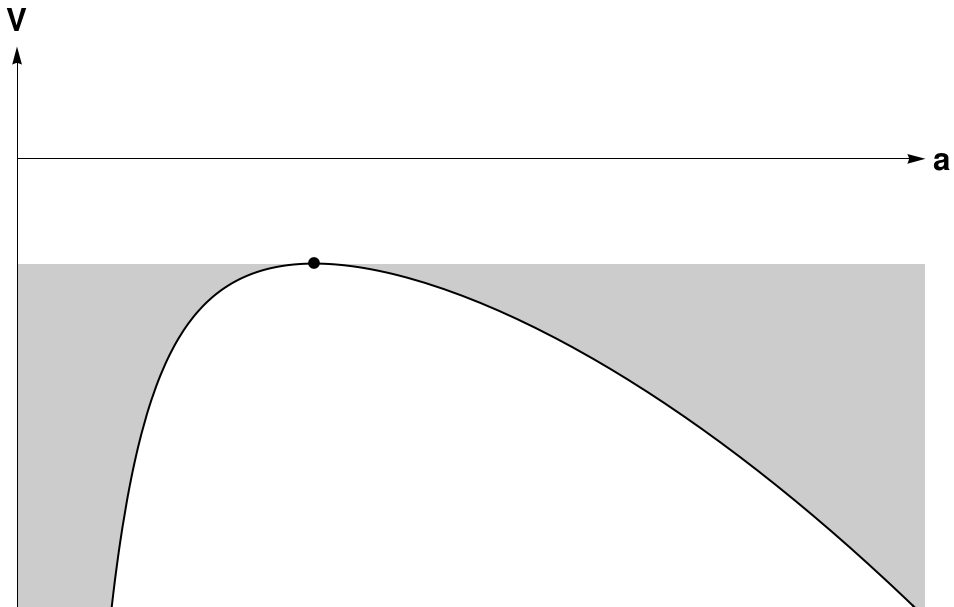}
\caption{Effective potential for scale factor with unstable equilibrium point.}
\label{fig: figure2}
\end{figure}

Because the metric in this model is similar to that of the closed FLRW universe, we end up with a system analogous to the Friedmann equations with the additional terms from electromagnetic stress-energy tensor,
\begin{equation} \label{eq: freq}
\begin{cases} 
3\left(\frac{1+\dot{a}^2}{a^2} \right) - \Lambda = 4\pi \frac{{E_0}^2}{a^4} + 8 \pi \rho, \\
-2\frac{\ddot{a}}{a} - \left(\frac{1+\dot{a}^2}{a^2} \right) + \Lambda = - 4\pi \frac{{E_0}^2}{a^4} + 8 \pi p,\\
-2\frac{\ddot{a}}{a} - \left(\frac{1+\dot{a}^2}{a^2} \right) + \Lambda = 4\pi \frac{{E_0}^2}{a^4}.
 \end{cases}
\end{equation}
As it turns out, we can deal with system (\ref{eq: freq}) using a similar procedure as in the standard case \cite{MTW}. The first equation is equivalent to
\begin{equation} \label{eq: fried1}
\frac{a}{2} + \frac{a \dot{a}^2}{2} - \frac{\Lambda a^3}{6} - \frac{2 \pi}{3} \frac{{E_0}^2}{a} = \frac{4 \pi}{3} \rho a^3.
\end{equation}
After taking a time derivative of both sides and using the third equation from (\ref{eq: freq}), we find
\begin{equation} \label{eq: friedrho}
 \rho =\frac{3 M}{4 \pi a^3} + \frac{{E_0}^2}{a^4},
\end{equation}
where $M$ is another integration constant. We can use now the second equation from (\ref{eq: freq}) to obtain the pressure $p$ as
\begin{equation} \label{eq: friedp}
p = \frac{{E_0}^2}{a^4}.
\end{equation}
Note that the matter satisfies the weak, strong and dominant energy conditions if $M \geq 0$, as in this case $0 < p \leq \rho$. Nevertheless,  $p/\rho$ always approaches $1$ as $a$ tends to zero, indicating that the matter particles approach the speed of light near the Big Bang or the Big Crunch. In the limit case $M=0$ we have $p=\rho$, meaning that the particles are actually moving at the speed of light.
\\
\indent
The behavior of the scale factor $a$ can be derived from an effective potential formulation. Using (\ref{eq: fried1}) and (\ref{eq: friedrho}) we have 
\begin{equation}
\frac{1}{2}\dot{a}^2 + \left( - \frac{\Lambda a^2}{6} - \frac{2\pi{E_0}^2}{a^2} - \frac{M}{a}   \right) = -\frac{1}{2}.
\end{equation}
We can think about each term as a kinetic part, an effective potential $V(a)$, and a conserved energy $h=-1/2$, respectively,
\begin{equation} \label{eq: effpot}
\frac{1}{2}\dot{a}^2 + V(a) = h.
\end{equation}
The potential is a function of $a$, but also depends on the three parameters $\Lambda$, $M$ and $E_0$. Clearly we need a positive cosmological constant to have an (unstable) equilibrium point, so that we can  reproduce the previous results concerning the Einstein universe. In this case, $V(a)$ will have the shape depicted on Fig.~\ref{fig: figure2}. Depending on the values of the parameters, we can have expanding solutions, recollapsing solutions, de Sitter-like bouncing solutions, and of course the unstable static solution and its asymptotes. Note that in units for which the radius of the static solution is $a=1$, we have
\[
\begin{cases}
\Lambda = 1 + 4\pi {E_0}^2, \\
M = \frac13(1 - 8\pi {E_0}^2)
\end{cases}
\]
(hence we must have ${E_0}^2 \leq \frac{1}{8\pi}$).

Finally, we note that except in the case of the Einstein universe and its two asymptotes, the electromagnetic field performs a finite number $n$ of oscillations during the history of the universe, proportional to its lifetime as measured in conformal time:
\[
n = \frac1{\pi} \int_0^{+\infty}\frac{dt}{a}.
\]
Therefore, at any given epoch these fields will appear mostly constant and may perhaps serve as models of primordial magnetic fields \cite{Thorne}.

% \section{Discussion}
% This paper explores only a part of existing solutions of the Maxwell field equations on the Einstein universe. In order to find the existence of more of them, an ansatz concerning the most general form of fields $E$ and $B$ has to be made. Unfortunately in this case the system (\ref{eq:MaE}) becomes complicated and no decoupling of equations is possible. 
% \\
% \indent
% The stress-energy tensor of the form (\ref{eq: setmex}) can be thought of as a sum two components representing pressureless fluids with the same mass density moving in opposite directions along $X_1$. Another interpretation, both in the static and expanding universe, is given by the cosmic strings of matter aligned with the Clifford parallels, expanding and contracting with the space and indicating the structure of Hopf fibration.


\begin{thebibliography}{1}

%\bibitem{bonola}
%R.\ Bonola, {\em Non-Euclidean geometry}, Dover, 2012.
\bibitem{MTW}
C.\ Misner, K.\ Thorne and J.\ Wheeler, {\em Gravitation}, Freeman, 1973.
\bibitem{wald}
R.\ Wald, \emph{General relativity}, University of Chicago Press, 1984.
\bibitem{HE95}
S.\ Hawking and G.\ Ellis, \emph{The large scale structure of space-time},
  Cambridge University Press, 1995.
\bibitem{Bicak1}
J.\ Bicak and P.\ Krtous, \emph{The fields of uniformly accelerated charges in de Sitter spacetime}, Phys.Rev.Lett. {\bf 88} (2002) 211101
\bibitem{Bicak2}
J.\ Bicak and P.\ Krtous, \emph{Fields of accelerated sources: Born in de Sitter}, J.Math.Phys. {\bf 46} (2005) 102504
\bibitem{ranada2}
A.\ Ra\~nada, {\em Topological Electromagnetism}, J.\ Phys.\ A: Math.\ Gen.\ {\bf 25} (1992) 1621-1641.
\bibitem{ranada1}
A.\ Ra\~nada, {\em Knotted solutions of the Maxwell equations in vacuum}, J.\ Phys.\ A: Math.\ Gen.\ {\bf 23} (1990) L815-L820.
\bibitem{ranada3}
A.\ Ra\~nada and J.\ Trueba, {\em Electromagnetic knots}, Phys.\ Lett.\ A {\bf 202} (1995) 337-342.
\bibitem{Thorne}
K.\ Thorne, {\em Primordial Element Formation, Primordial Magnetic Fields, and the Isotropy of the Universe}, Astrophys.\ J.\ {\bf 148} (1967) 51-68.
\end{thebibliography}
\end{document}